\documentclass[%
,preprint%
,secnumarabic
,amssymb]{revtex4}
\usepackage{amsmath}%
\usepackage{bm}%
\usepackage{epsf}%
\usepackage{ifpdf}%

\ifpdf
\usepackage[pdftex]{graphicx}
\else
\usepackage{graphicx}
\fi

\ifpdf
\DeclareGraphicsExtensions{.pdf, .jpg}
\else
\DeclareGraphicsExtensions{.eps}
\fi

\begin{document}
\title{Analysis of  Reflection Electron Energy Loss Spectra (REELS) \\
for Determination of the  Dielectric Function of Solids: Fe, Co, Ni.}

\author{ Wolfgang S.M. Werner\footnote{werner@iap.tuwien.ac.at, \\
fax:+43-1-58801-13499,~tel:+43-1-58801-13462}}%
\affiliation{Institut f{\" u}r Allgemeine Physik,~Vienna University of Technology,
Wiedner Hauptstra\ss e 8--10,~A~1040~Vienna,~Austria
}
\date{\today}%
\begin{abstract} 
A simple procedure is developed to simultaneously eliminate multiple scattering contributions from  two reflection
electron energy loss spectra (REELS) measured at different energies or for different experimental geometrical
configurations.   The procedure provides the differential inverse inelastic mean free path (DIIMFP) and the
differential surface excitation probability (DSEP).  The only required input parameters are the differential cross
section for elastic scattering and a reasonable estimate for the inelastic mean free path (IMFP). No prior information on
surface excitations is required for the deconvolution. The retrieved DIIMFP and DSEP  can be used to determine the
dielectric function of a solid by fitting the DSEP and DIIMFP to theory. Eventually, the optical data can be used to
calculate the (differential and total) inelastic mean free path and the surface excitation probability.
The procedure is applied to Fe, Co and Ni and the retrieved optical data as well as the inelastic mean free paths and
surface excitation parameters derived from it are compared to values reported earlier in the literature. In all cases,
reasonable agreement is found between the present data and the earlier results, supporting the validity of the
procedure.
\newline
PACS numbers: 68.49.Jk, 79.20.-m, 79.60.-i
\end{abstract}

\maketitle 
\section{Introduction.}
The response of a solid to an external electromagnetic perturbation is decribed by the dielectric function
$\varepsilon(\omega,q)$, where $\hbar\omega$ and $\hbar q$ are the energy and momentum transfered during the
interaction. Knowledge of the dielectric function of a solid is important  for many branches of physics. 
The dielectric function  can be measured by probing a solid surface with elementary particles,
 e.g. by photons \cite{palik,palik1,henke,chantler,chantler1} or electrons \cite{schattschneider}.
 With the advent of density functional theory beyond the ground state
\cite{kohndft}, {\em  ab initio} theoretical calculations of optical data have recently become available
\cite{wien2kcpc,weroptcu,wersub}. 

From the experimental point of view, reflection electon energy loss experiments represent a particularly attractive
method to probe the dielectric response of a solid, since the experiment is very simple. However, quantitative
interpretation of the experimental results is not straightforward since inside the solid, the electrons experience
intensive multiple scattering implying that experimental data need to be deconvoluted before information on the
dielectric response of the solid  can be extracted from them. In the past, the algorithm of Tougaard and
Chorkendorff  (hereafter designated by "TC") \cite{toupr35} has been frequently used for this purpose. However, it has
recently been shown \cite{werbivar,werbivarss} that the loss distribution retrieved by the TC-algorithm is not unique,
i.e., it constitutes a not very well defined mixture of so--called surface and volume electronic excitations of any
scattering order, which makes quantitative interpretation and extraction of the dielectric function troublesome.   An
algorithm which is free of these deficiencies has been proposed recently by the present author (denoted as "W"-algorithm
hereafter) and was succesfully applied to several materials \cite{werbivar,werbivarss,weroptcu,wersub}.  The W-algorithm
is more involved than the TC-algorithm and needs evaluation of many terms for which the expansion coefficients are
tedious and time consuming to obtain. In the present work, a simplification of the W-algorithm, (denoted by
"SW"-algorithm) is developed, following a suggestion made earlier in this connection by Vicanek \cite{vicanek} and
succesfully tested using REELS spectra of Fe, Co and Ni. The resulting optical data and quantities derived from them
such as the mean free path for inelastic electron scattering  are in good agreement with data found in the
literature, supporting the validity of the procedure.

\section{Deconvolution of REELS Spectra.}
 A REELS spectrum is made up of electrons that have experienced surface (designated by the subscript "s" in the
following) and bulk (subscript "b") excitations  a certain number of times. The number of electrons reaching the
detector after participating in $n_{s}$ surface and $n_{b}$ bulk excitations is given  by the partial intensities
$A_{n_{b}n_{s}}$. Since the occurence of surface excitations is localized to a depth region smaller than, or comparable
to, the elastic mean free path, the partial intensities for surface and bulk scattering are uncorrelated to a good
approximation\cite{wernrelprl}:
\begin{equation}
\label{epiuncorr} A_{n_{b}n_{s}}=A_{n_{b}} \times A_{n_{s}}
\end{equation}
The bulk partial intensities can be obtained most conveniently by means of a  Monte Carlo calculation, by calculating
the distribution of pathlengths $Q(s)$ the electrons travel in the solid and using the formula \cite{wernist}:
\begin{equation}
\label{epld}
A_{n_{b}}=\int Q(s) W_{n_{b}}(s) ds
\end{equation}
where $W_{n}(s)$ is the stochastic process for multiple scattering:
\begin{equation}
\label{epoisson}
W_{n}(s)=(\frac{s}{\lambda_{i}})^n \frac{\exp(-s/\lambda_{i})}{n!}
\end{equation}
and $\lambda_{i}$ is the inelastic mean free path. The only physical quantity needed for the calculation of the
pathlength distribution is the differential cross section for elastic scattering \cite{jabsalpow}, which can be
calculated ab initio for free atoms if solid state effects are weak for the considered application. The pathlength
distribution for a reflection geometry is always much broader than the distribution $W_{n_{b}}(s)$ for any value of
$n_{b}$, implying that the reduced partial intensities $\alpha_{n_{b}}=A_{n_{b}}/A_{n_{b}=0}$ depend only very weakly on
the value of the inelastic mean free path.

 The surface partial intensities are also assumed to be governed by  Poisson
statistics. Since an electron's trajectory through the surface scattering zone is rectilinear to a good approximation
(i.e. the pathlength distribution strongly resembles a delta-function), the reduced partial intensities for surface
scattering  $\alpha_{n_{s}}=A_{n_{s}}/A_{n_{s}=0}$ are given by the simple equation \cite{werbivar,werbivarss}:
\begin{equation}
\label{ecns} \alpha_{n_{s}}=\langle n_{s} \rangle^{n_{s}} /n_{s}!
\end{equation}
where $\langle n_{s}\rangle$ is the average number of surface excitations taking place during reflection, i.e. the
incoming and outgoing surface crossing is combined in Eqn.~(\ref{ecns})\cite{werbivar}. Commonly, an expression of the form:
\begin{equation}
\label{ens}
\langle n_{s}\rangle=\frac{a_{s}}{\sqrt{E}}(\frac{1}{\mu_{i}}+\frac{1}{\mu_{o}})
\end{equation}
is adopted for average number of surface excitations \cite{tungpr49} where $a_{s}$ is the so--called surface excitation
parameter, $E$ is the electron energy and $\mu_{i}=\cos\theta_{i}$ and $\mu_{o}=\cos\theta_{o}$ are  the polar
directions of surface crossing for the incoming and outgoing beam.

Multiplying the number of electrons arriving in the detector $\alpha_{n_{b}n_{s}}$ with the energy distribution after
experiencing $(n_{b},n_{s})$ collisions, the partial loss distributions, $\Gamma_{n_{b}n_{s}}(T)=w_{b}^{(n_{b})}(T-T')\otimes
w_{b}^{(n_{b})}(T)$, and summing over all scattering orders, the energy loss  spectrum $y(T)$ is obtained as
\cite{werbivar}:
\begin{equation}
\label{ereels}
y(T)=\sum\limits_{n_{s}=0}^{\infty} \sum\limits_{n_{b}=0}^{\infty} \alpha_{n_{b}n_{s}}w_{b}^{(n_{b})}(T-T')\otimes
w_{s}^{(n_{s})}(T')
\end{equation}
Where $T$ denotes the energy loss, the symbol "$\otimes$" represents a convolution over the energy loss variable and the
quantities $w_{b}^{(n_{b})}$ and $w_{s}^{(n_{s})}$ are the $(n_{b}-1)$--fold and $(n_{s}-1)$--fold selfconvolution of
the normalized energy loss distribution in a single bulk and surface excitation respectively, the so--called
differential inverse inelastic mean free path (DIIMFP, $w_{b}(T)$) and differential surface excitation probability
(DSEP, $w_{s}(T)$). Note that the elastic peak needs to be removed from a REELS spectrum before analysis, implying that
$\alpha_{0,0}=0$ \cite{werbivar}.

The subject of the present paper is the retrieval of the quantities $w_{b}(T)$ and $w_{s}(T)$ from experimental REELS
spectra and determination of the dielectric function from these quantities. Recalling the convolution theorem, it is
immediately obvious that in Fourier space, the spectrum is given by a bivariate power series in the variables  $w_{b}(T)$ and
$w_{s}(T)$. This implies that a unique solution of these quantities cannot be found by reverting the series
Eqn.~(\ref{ereels}) since a single equation with two unknowns has no unique solution. However, when {\em two } loss spectra
 ${y}_{1}(T)$ and  ${y}_{2}(T)$ with a
different sequence of  partial intensities  $\alpha_{n_{b}n_{s}}$ and $\beta_{n_{b},n_{s}}$, are measured, reversion
of the bivariate power series becomes possible using the formulae \cite{werbivar,werbivarss}:
\begin{eqnarray}
\label{eformalreversion}
{w}_b(T)
&=&
\sum\limits_{p=0}^\infty
\sum\limits_{q=0}^\infty
u_{p,q}^b 
{y}_{1}^{(p)}(T-T')\otimes
{y}_{2}^{(q)}(T')
\nonumber\\
{w}_s(T)
&=&
\sum\limits_{p=0}^\infty
\sum\limits_{q=0}^\infty
u_{p,q}^s 
{y}_{1}^{(p)}(T-T')\otimes
{y}_{2}^{(q)}(T'),
\end{eqnarray}
The above expression constitutes the  W-algorithm for which the required coefficients $u_{p,q}^b$ and
$u^s_{p,q}$ can be obtained as outlined in Refs.~\cite{werbivar,werbivarss}. 

\begin{figure}[htb]
{\includegraphics[width=11.0cm]{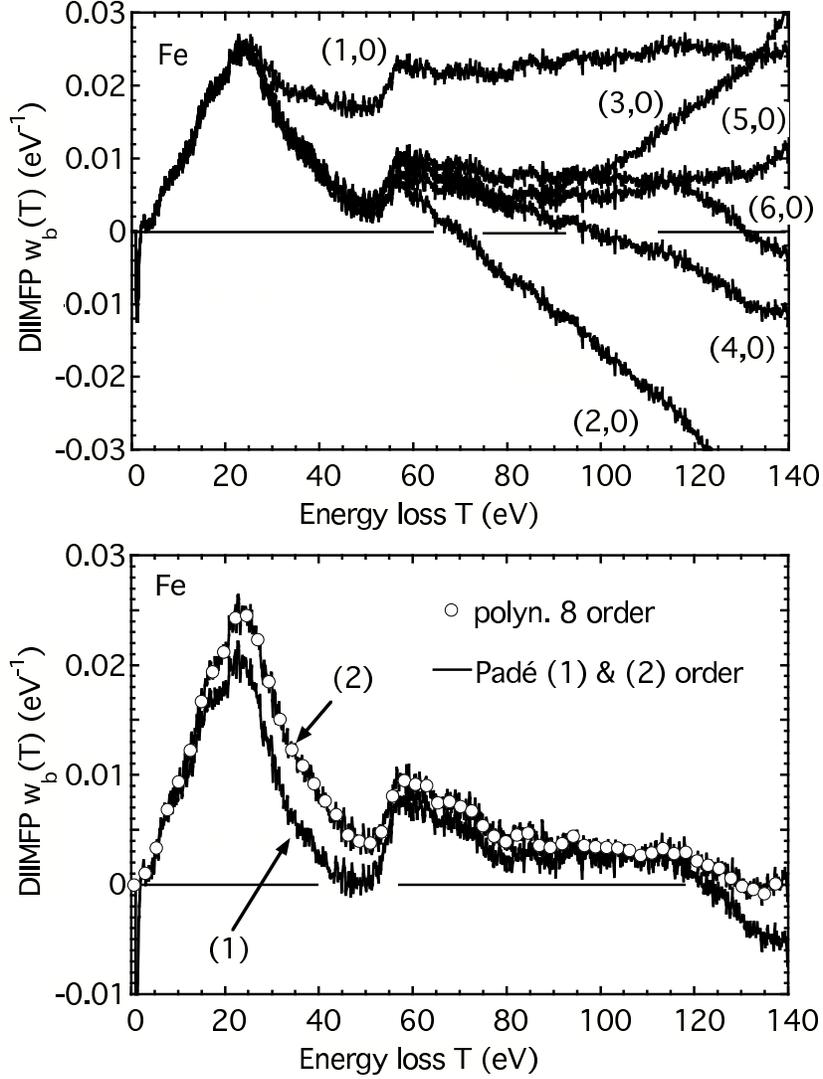}}
\caption{%
Illustration of the successive stages of the deconvolution of REELS spectra of Fe (measured at 1000 and 3400~eV) using
an eighth order polynomial approximation (a.) and a second and first order Pad\'e rational fraction approximation
(Eqn.~(\ref{epade})) (b). The open circles in (b)   are the result of the 8--th order  polynomial expansion
Eqn.~(\ref{eformalreversion}), the solid curves labelled (1) and (2) are the first and second order Pad\'e
approximation, respectively. The curves shown represent the DIIMFP, the corresponding results for the DSEP are not shown
for clarity, but behave in a similar way.}
\label{fpolpadfe}
\end{figure}

The W-algorithm has been succesfully applied to REELS data of a large number of materials
\cite{werbivar,werbivarss,zemwerecasia05} and the dielectric function of several materials was succesfully extracted from
the resulting DIIMFP and DSEP \cite{weroptcu,wersub}. However, this approach may be improved upon in two respects:
first of all, convergence of the series Eqn.~(\ref{eformalreversion}) is relatively slow, implying that many terms
need to be calculated; secondly, calculation of the higher order coefficients $u_{p,q}$ is tedious and time
consuming.
The moderate convergence behaviour of the  W-algorithm is illustrated in Figure~\ref{fpolpadfe}a that shows the
various stages of the deconvolution. The first collision order $n$ indicated for the different curves implies
that the second collision order has been taken into account up to $n-1$, so, for example,  the curve labelled $(3,0)$ is
a superposition of the $(1,0), (0,1), (1,1), (2,0), (2,1), (2,2)$ and  $(3,0)$--th order cross-convolutions of the two
experimental spectra. It is seen that more than 6 scattering orders need to be taken into account for the W-algorithm to
converge over a loss range of about 100~eV.

The improvements addressed above can be effected following a suggestion by Vicanek\cite{vicanek} who pointed out 
that the TC-algorithm (for the univariate case)  can in fact be regarded as a lowest order rational fraction expansion
(a so--called Pad\'e approximation) of the power series expansion of the spectrum. Extension of the concept of the
Pad\'e approximation to multivariate power series, such as Eqn.~(\ref{eret}) is straightforward \cite{cuyt}.

The simplified  algorithm (designated by "SW"--algorithm hereafter) is obtained (in Fourier space, indicated by the ("$\widetilde{~}$")--symbol)
by making the rational fraction ansatz :
\begin{equation}
\label{epade}
\widetilde{w}=
\frac{
\sum\limits_{k=0}^{N}
\sum\limits_{l=0}^{N}
a_{k,l}
 \widetilde{y}^k_{1}
 \widetilde{y}^l_{2}
 }{
\sum\limits_{k=0}^{N}
\sum\limits_{l=0}^{N}
b_{k,l}
 \widetilde{y}^k_{1}
 \widetilde{y}^l_{2}
 }
\end{equation}
where $N$ is the order of the approximation and with $b_{0,0}=1$ and $a_{0,0}=0$. Note that the same formula holds for
the quantities
$w_{s}(T)$ and $w_{b}(T)$, only the expansion coefficients are different. Explicit expressions for the first order  Pad\'e
coefficients as well as  the system of linear equations determining the second order expansion coefficients, are given
in the appendix. Multiplying Eqn.~(\ref{epade}) with the denominator of the right hand side, and going back to real
space one finds the deconvolution formula to be given by a Volterra integral equation of the second kind (being most
harmless numerically):
\begin{equation}
\label{eret}
w(T)=\sum\limits_{k=0}^{N}\sum\limits_{l=0}^{N} a_{k,l}Y_{k,l}(T)
-\int\limits_{T'=0}^T 
\sum\limits_{k=0}^{N}\sum\limits_{l=0}^{N} b_{k,l}Y_{k,l}(T-T')w(T')dT'
\end{equation}
where the quantity $Y_{k,l}(T)$ is the $(k,l)$-th order cross convolution of the two REELS spectra:
\begin{equation}
\label{ecrossconv}
Y_{k,l}(T)=y_{1}^{(k)}(T-T')\otimes y_{2}^{(l)}(T')
\end{equation}
Since $w(T=0)\equiv 0$ the integration on the right hand side of Eqn.~(\ref{eret}) can always be carried out over the
energy loss range for which the loss distribution is already known. For more details on the (trivial) numerical
treatment of the Volterra integral equation of the second kind see, e.g., Ref.~\cite{press}.

The performance of the SW-algorithm is illustrated in Figure~\ref{fpolpadfe}b, that shows
the 8--th order polynomial expansion approximation (Eqn.~(\ref{eformalreversion})) as open circles. The solid curves
labelled (1) and (2) are the first and second order Pad\'e approximation results. It is seen that the latter converges
as good as the polynomial expansion over the first $\sim$150~eV. Even the first order approximation leads to reasonable
agreement. However, at least the second order mixed term (one surface and one bulk excitation) is important to attain
quantitative agreement. The reason for the good performance of a rational fraction approximation compared to 
the polynomial expansion is clear: for a polynomial expansion, for any value of the variables a certain power in
the series will always be the dominating term  and  may become arbitrarily large. For a rational fraction expansion, a
large value of the dominant term in the enumerator can always be balanced by an appropriately chosen coefficient  of the
power in question in the denominator.

The second order SW-algorithm was applied REELS spectra of  Ag, Al, Au, Be, Bi, C, Co, Fe, Ge, Mo, Mn, Ni, Pb, Pd, Pt,
Si, Ta, Te, Ti, V, W and Zn) and was found to converge over the measured energy loss range of 140~eV.  It can therefore
be concluded that for an energy loss range of this order the second order SW-algorithm is sufficiently accurate. The
main merit of the first order approximation is that the expansion coefficients can be given in a tractable analytical
form (see the Appendix), providing detailed insight in the physical behaviour of the deconvolution. For example, it is
seen in Eqn.~(\ref{epad1coef}) and (\ref{epad2coeff}) that the retrieved DIIMFP is completely independent of the value
of the surface excitation parameter $a_{s}$ used for the deconvolution. On the other hand, the expansion coefficients
for the surface loss probability are seen to scale linearly with $a_{s}$.  This can be seen by substituting
Eqn.~(\ref{epiuncorr}) in Eqn.~(\ref{epad1coef}) and (\ref{epad2coeff}) and using Eqn.~(\ref{ecns}) and
Eqn.~(\ref{ens}). Thus, no information on surface excitations whatsoever is needed as input to the procedure (except the
functional from of the dependence of the surface excitation probability on the energy and surface crossing direction,
such as Eqn.~(\ref{ens})). Furthermore, as pointed out above,  the value of the inelastic mean free path  will mainly
affect the absolute value of the partial intensities, $A_{n_{b}}$, while the reduced partial intensities
$\alpha_{n_{b}}$ change negligibly when the IMFP is varied by up to 30\%. Reasonably accurate knowledge of the shape and
magnitude of the elastic scattering cross section is important, however, since the shape of the pathlength distribution
depends on it. Near a deep minimum of the cross section this dependence may even be critical
\cite{wernist}, giving  rise to qualitatively different sequences of partial intensities. 

The higher order expansion coefficients also exhibit the scaling properties addressed above. Therefore, using a set of
input parameters as discussed above, the normalized DIIMFP is correctly returned by the procedure 
in absolute units, while the correct shape of the DSEP is also obtained, but the absolute value of the DSEP scales with
the value of $a_{s}$ used in the procedure.

Finally, it should be pointed out that by a suitable choice of the partial intensities (in particular letting
$\beta_{n_{b}n_{s}}\rightarrow 0$, mimicking the use of a single REELS spectrum) Eqn.~(\ref{eret}) reduces to the
algorithms proposed by Vicanek \cite{vicanek}  or Tougaard and Chorkendorff  \cite{toupr35}.

\section{Retrieval of Optical Data from the Single Scattering Distributions.}
The procedure to extract the dielectric function of a solid from the DSEP and the DIIMFP has been outlined in
Ref.~\cite{weroptcu}: The theoretical expressions for the DIIMFP and DSEP are fitted to the corresponding experimental
results using a suitable model for the dielectric function.
The differential inverse inelastic mean free path is related to the dielectric function $\varepsilon( \omega,q)$ of the
solid via the well known formula \cite{landauem}:
\begin{equation}
\label{ediimfp}
W_b(\omega)=\frac{1}{\pi E}\int\limits_{q_-}^{q_+} {\cal I}m\Big\{ \frac{-1}{\varepsilon(
\omega,q)}\Big\}\frac{dq}{q}.
\end{equation}
For parabolic bands, the momentum transfer $q$
confined by $q_-$ and $q_+$ given by:
\begin{equation}
\label{qminmax}
q_\pm=\sqrt{2E}\pm\sqrt{2(E-\omega)}.
\end{equation}
Note that atomic units are used in this section. The  differential inverse mean free path
normalized to unity area is related to the unnormalized DIIMFP via
$w_b(T)=\lambda_i W_b(T)$. 

For the DSEP, the expression by Tung and coworkers \cite{tungpr49} is used:
\begin{equation}
\label{etung}
W_s(\omega,\theta,E)=P_s^+(\omega,\theta,E)+P_s^-(\omega,\theta,E),
\end{equation}
where the quantity $P_s^\pm(\omega,\theta,E)$ is defined as
\begin{equation}
\label{edsep}
P_s^\pm(\omega,\theta,E)=\frac{1}{\pi E\cos\theta}\int\limits_{q_-}^{q_+}\frac{|q_s^\pm |dq}{q^3}
{\cal I}m\Big[\frac{(\varepsilon(\omega,q)-1)^2}{\varepsilon(\omega,q)(\varepsilon(\omega,q)+1)}\Big],
\end{equation}
and 
\begin{equation}
\label{etung3}
q_s^\pm =\Big[q^2-\big(\frac{\omega+q^2/2}{\sqrt{2E}}\big)^2\Big]^{1/2}\cos\theta\pm
(\frac{\omega+q^2/2}{\sqrt{2E}})\sin\theta.
\end{equation}
The normalized differential surface excitation probability is calculated via  $w_s(T,\theta)=W_s(T,\theta)/\langle
n_s(\theta,E)\rangle$.

 The electromagnetic response of the solid is described in terms of the fit-parameters $f_{i},\omega_{i}$ and $\gamma_{i}$
 using a  model for the dielectric function in terms of a set of Drude-Lindhard oscillators for the evaluation
of the theoretical expressions for the DIIMFP and DSEP \cite{chenpr48}:
\begin{eqnarray}
\label{edrude}
\varepsilon_{1}(\omega,q)&=&\varepsilon_{b}-\sum\limits_{i}\frac{f_{i}(\omega^2-\omega_{i}(q)^2)}{(\omega^2-%
\omega_{i}(q)^2)^2+\omega^2\gamma_{i}^{2}}\nonumber\\
\varepsilon_{2}(\omega,q)&=&\sum\limits_{i}\frac{f_{i}\gamma_{i}\omega}{(\omega^2-\omega_{i}(q)^2)^2+\omega^2%
\gamma_{i}^{2}},
\end{eqnarray}
where $\omega=T$ is the energy loss and $q$ is the momentum transfer. The static dielectric constant is given by 
$\varepsilon_{b}$, $f_{i}$ represents the oscillator strength, $\gamma_{i}$ the damping coefficient and $\omega_{i}$ is
the energy of the $i$--th oscillator. Those values of the Drude--Lorentz parameters in the above equation that minimize
the deviation between the experimental data and the theoretical expressions for the DIIMFP and DSEP are taken to
parametrize the dielectric function. A quadratic dispersion  $\omega_{i}(q)=\omega_{i}+q^2/2$ was used for the
transition described by the $i$--th oscillator, except for those oscillators with an energy higher than the most loosely
bound core electrons, that are assumed to be described by localized states without dispersion.  Note that a dielectric
function of the form Eqn.~(\ref{edrude}) implicitly satisfies the Kramers-Kronig dispersion relationships. 

The (normalized) distributions $w_{b}(T)$ and $w_{s}(T)$ obtained from the experimental data are simultaneously fitted to
the above theoretical expressions by minimizing the function:
\begin{equation}
\label{echisquare}
X_{b}\chi^2_{DIIMFP}(p_{b}, f_{i},\omega_{i},\gamma_{i})+X_{s}\chi^2_{DSEP}(p_{s}, f_{i},\omega_{i},\gamma_{i}).
\end{equation}
Here the function $\chi^2$ is the least squares difference between theory and experiment, defined in the usual way and
$X_{b}$ and $X_{s}$ are weight factors chosen as  $X_{b}=1.0$ and $X_{s}=0.1$, emphasizing the DIIMFP in the fitting
procedure. The fit-parameters $p_{b}$ and $p_{s}$ are multiplicative scaling factors for the DIIMFP and the DSEP that 
compensate an eventual mismatch of the surface excitation parameter and IMFP used as input. Using the TPP-2M formula
\cite{tansia21} to estimate the IMFP, the value of $p_{b}$ deviated from unity by less than 5\% for all cases studied,
while the returned value of $p_{s}$ scales with the value  of the surface excitation parameter $a_{s}$, exactly as
anticipated in the previous section.

\section{Experimental}
The procedure used to acquire the experimental data  has been described in detail before \cite{werepesjes}. REELS data
were taken for polycrystalline  Fe, Co and Ni surfaces  in the energy range between 300 and 3400~eV for normal incidence
and an off-normal emission angle of 60$^\circ$, using a hemispherical analyzer operated in the constant-analyzer-energy
mode giving a width of the elastic peak of 0.7~eV. Count rates in the elastic peaks were kept well below the saturation
count rate of the channeltrons  and a dead time correction was applied to the data. The sample was cleaned by means of
sputtering with 3~keV Ar$^+$ ions. Sample cleanliness was monitored with Auger electron spectroscopy.

\begin{figure}[htb]
{\includegraphics[width=11.0cm]{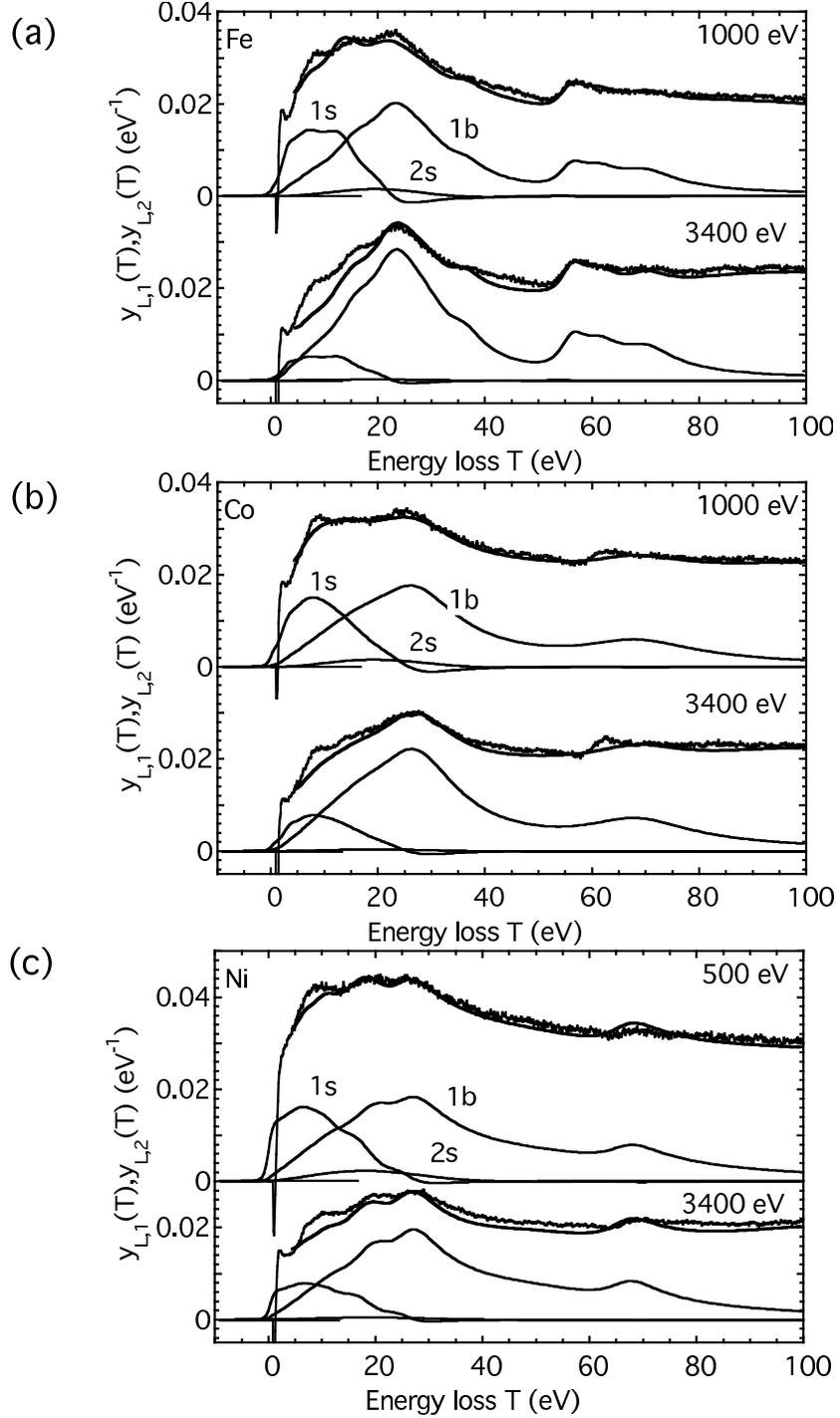}}
\caption{%
Experimental loss spectra for Fe, Co and Ni used in the present work (noisy curves). The thick solid curve is a fit of
these data to Eqn.~(\ref{ereels}), using the retrieved optical data to describe the DIIMFP and DSEP. 
The thin solid
curves labelled "1s", "1b" and "2s" represent the contribution of electrons having experienced one surface, one bulk and
one and two surface excitations respectively.
}
\label{freels}
\end{figure}

For each material the optimum energy combination for the retrieval procedure of 
two loss spectra was determined by inspection of the partial intensities, choosing those energies for which on the one
hand the determinant $\Delta$ in Eqn.~(\ref{edet}) is reasonably large, while, on the other hand, those energies were
avoided for which
the scattering geometry corresponds to a deep minimum in the elastic cross section.  For these cases it is more
difficult to obtain the realistic shape of the pathlength distribution since the elastic cross sections are not
accurately known for such scattering angles and the true electron optical detector solid angle also plays a significant
role there. In Figure~\ref{freels}, the experimental spectra used in the present work are shown as noisy curves. For all
selected energy combinations, the shape of the loss spectra is seen to be quite similar, but in all cases, a significant
difference in the relative contribution of surface and bulk excitations is seen, as evidenced by the difference in
spectral shape below $\sim$20~eV.

Removal of the elastic peak was achieved by fitting the elastic peak to a combination of a Gaussian and a Lorentzian
peakshape. Subsequently, the fitted elastic peak was subtracted from the experimental data, they  were divided by the
area of the elastic peak,  and the energy scale was converted to an energy-loss scale. Finally, the measured spectrum
$S(T)$ [in counts per channel] was converted to experimental yield $y(T)$ [in reciprocal eV],  corresponding to
Equation~(\ref{ereels}), by division by the channel width $\Delta E$. Note that due to the dynamical range of a typical
REELS spectrum measured with good energy resolution, a small misfit of the tail of the elastic peak can give erratic
excursions in the loss spectra obtained in this way. This can be observed in Figure~\ref{freels} for energy losses below
$\sim$5~eV, where a negative excursion and a small shoulder right next to it are seen that are due to elimination of the
elastic peak. The only way to cure this problem is to conduct the experiment with a better energy resolution, implying
that the primary beam must be monochromatized.

\section{Results.}
\begin{figure}[htb]
{\includegraphics[width=11.0cm]{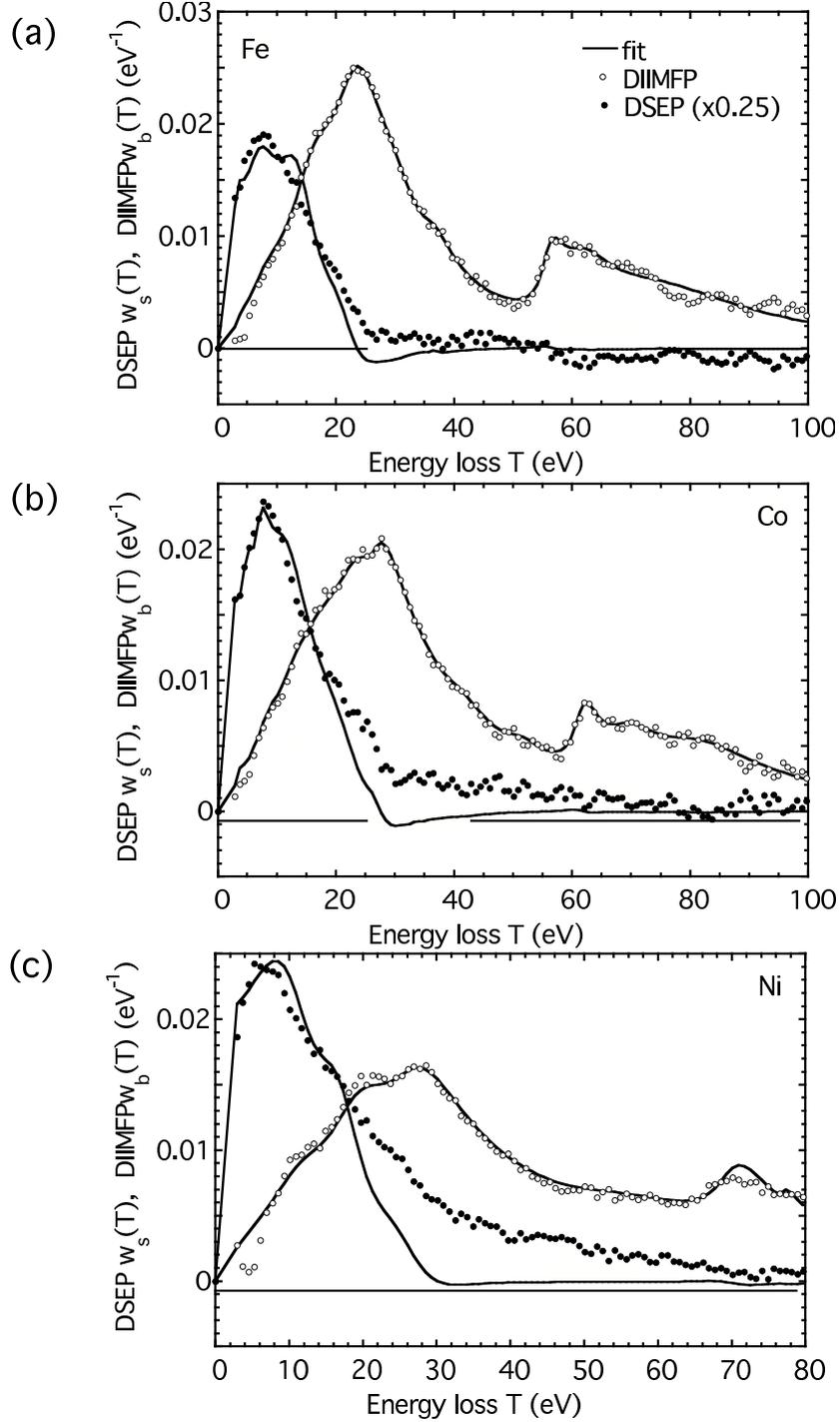}}
\caption{%
Normalized differential surface excitation probabililty (DSEP, $w_{s}(T)$) and normalized
differential inverse inelastic mean free path (DIIMFP, $w_{b}(T)$) retrieved from the experimental data by means of
Eqn.~(\ref{eret}). The solid line is obtained by simultaneously  fitting  these data to 
theory for a set of Drude-Lorentz fit--parameters for the dielectric function, as
described by Eqn.~(\ref{edrude}). The data for the surface excitation probability were scaled by a factor of 0.25 in
order to facilitate comparison. (a) Fe; (b) Co; (c)
 Ni.}
\label{ffit}
\end{figure}

The DIIMFP and DSEP retrieved from the spectra displayed in Figure~\ref{freels} with the second order SW--algorithm  are
presented in Figure~\ref{ffit} as open (DIIMFP) and filled circles (DSEP). The solid lines represent
the best fit of the data to theory. It is seen that the DIIMFP can be perfectly fitted by the employed theory, while
the corresponding DSEP agrees reasonably with the experimental data, but significant deviations are nonetheless
observed. This is believed to be attributable to the simplifications concerning the depth dependence of the surface
excitation process made in the employed theory for surface excitations \cite{tungpr49}. The Drude--Lorentz parameters
giving the best fit between experiment and theory are given in Table~1. The binding energy of the most loosely
bound core electrons \cite{sessa} are also indicated there and are in good agreement with the  energies of the ionization
edges observed in Figure~\ref{ffit}.

\begin{figure}[htb]
{\includegraphics[width=11.0cm]{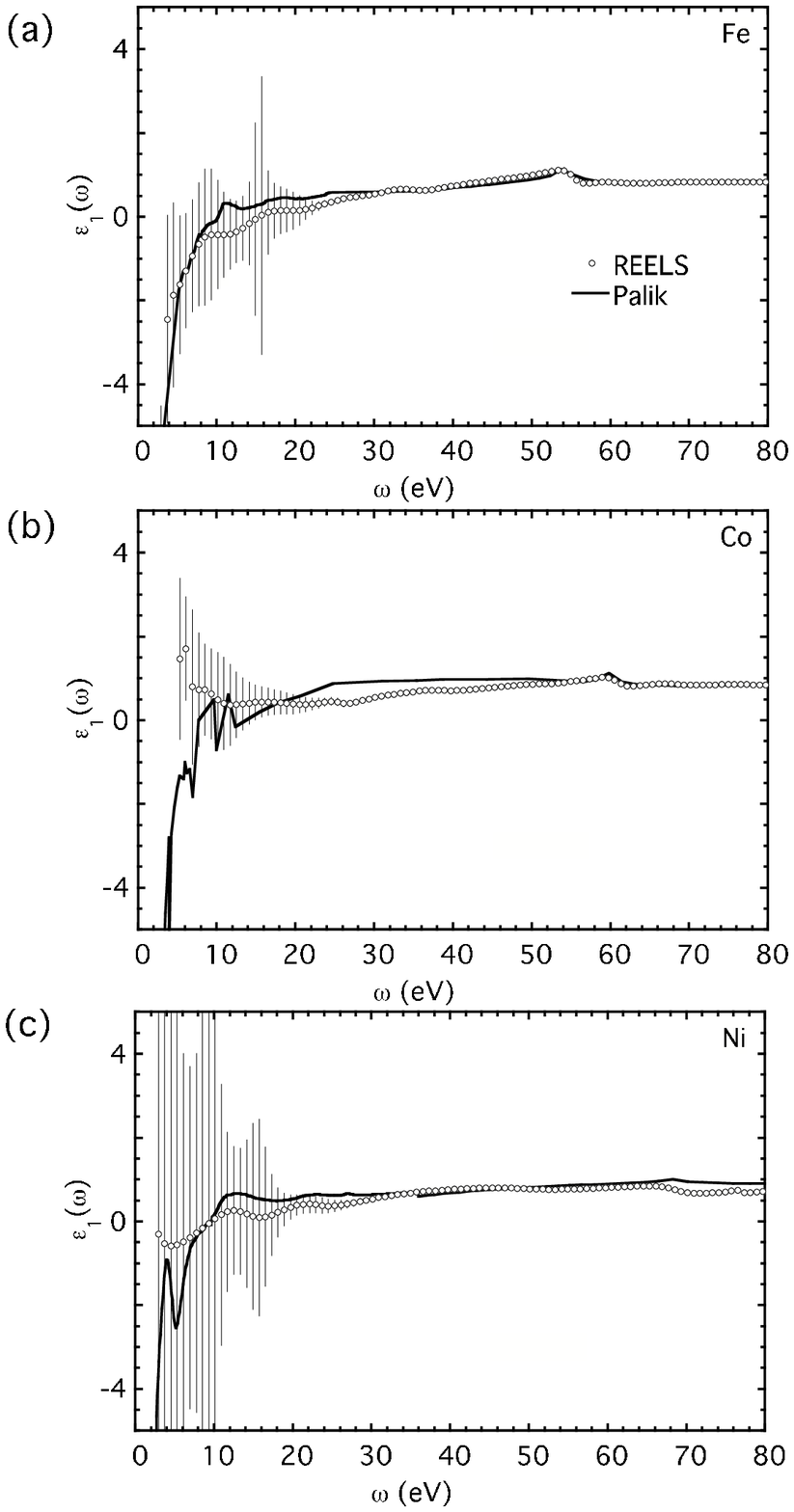}}
\caption{%
Real part of the dielectric function, $\varepsilon_{1}$,  of Fe, Co and Ni retrieved from the REELS data (open circles)
 compared with Palik's data \cite{palik,palik1} (solid curves). (a) Fe; (b) Co; (c)
 Ni.
}
\label{feps1}
\end{figure}

\begin{figure}[htb]
{\includegraphics[width=11.0cm]{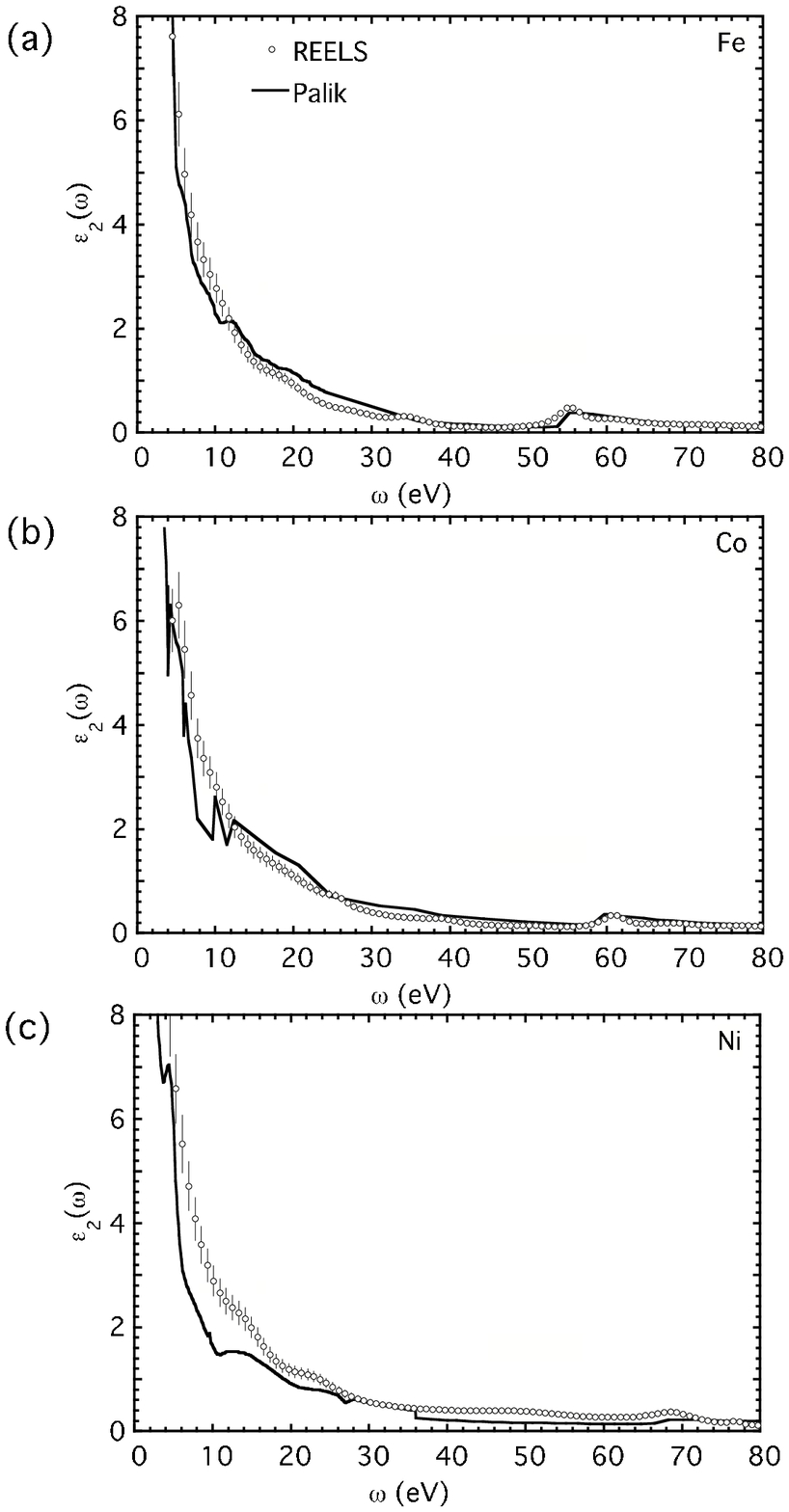}}
\caption{%
Imaginary part of the dielectric function, $\varepsilon_{1}$,  of Fe, Co and Ni retrieved from the REELS data (open
circles)  compared with Palik's data \cite{palik,palik1} (solid curves). (a) Fe; (b) Co; (c)
 Ni.
}
\label{feps2}
\end{figure}

A comparison of the real and imaginary part of the dielectric function derived from the REELS measurements with the
data given in Palik's book \cite{palik,palik1} is presented in Figure~\ref{feps1} and \ref{feps2}. Reasonable agreement
between these two data sets is observed for all cases for energies $\agt$5~eV. The error in the present data can
become excessively large below 5~eV  due to problems with the elimination of the elastic peak from the spectra. The
error bars in these graphs are obtained \cite{wersub} by assuming that the retrieved DIIMFP predominantly determines
$\varepsilon_{2}$ implying that the uncertainty in $\varepsilon_{2}$ is of the order of the uncertainty in the
retrieved DIIMFP.  The rules of error propagation are then used to estimate the uncertainty in $\varepsilon_{1}$. As can
be seen the uncertainty in $\varepsilon_{1}$ is rather large below 20~eV. This is a fundamental characteristic of the
derivation of optical data from absorption measurements, that mainly sample $\varepsilon_{2}$. Within the estimated
uncertainty, the two data sets agree satisfactorily, both for the real as well as the imaginary part of $\varepsilon$.

\begin{figure}[htb]
{\includegraphics[width=11.0cm]{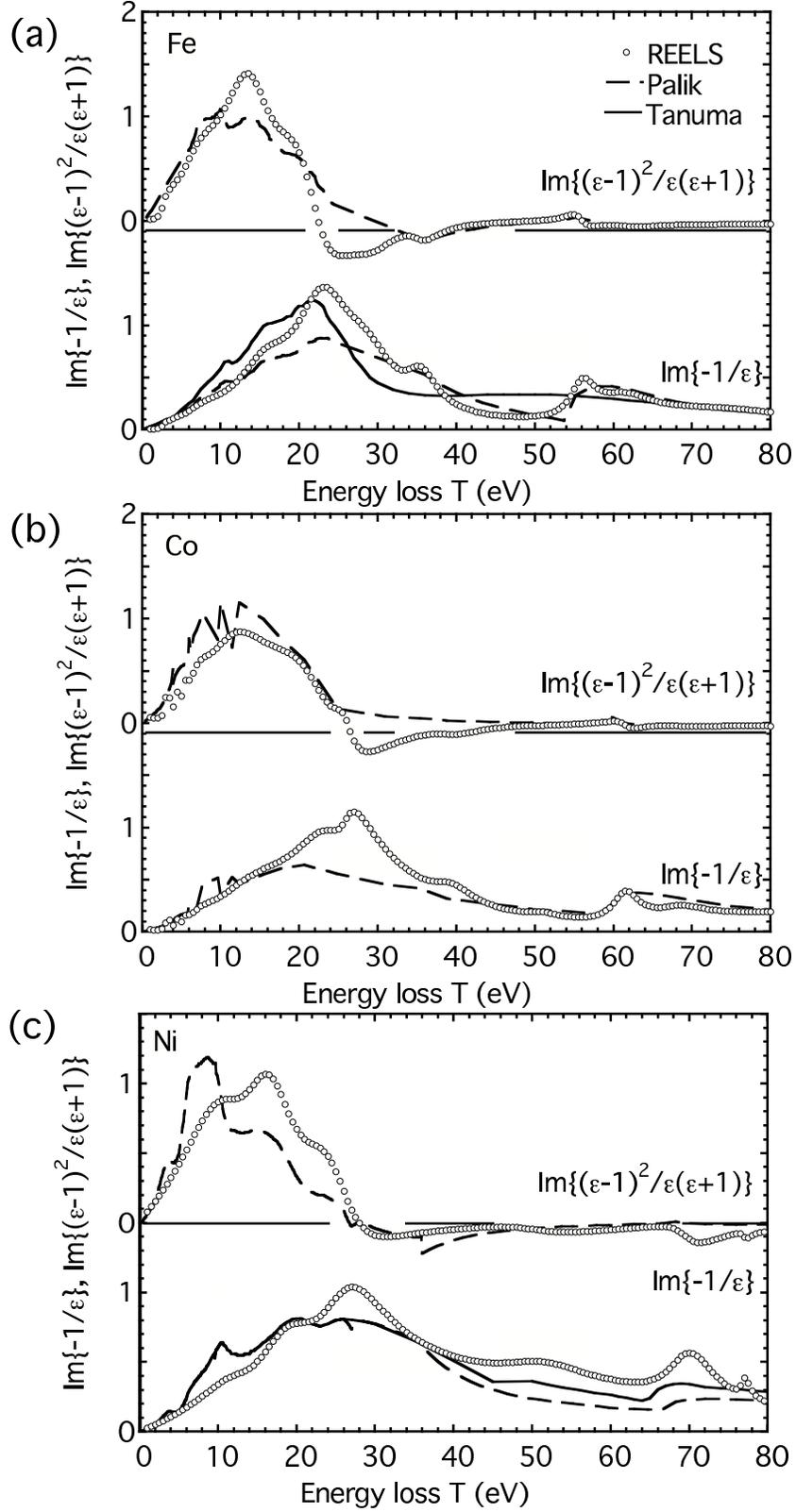}}
\caption{%
Volume  and surface loss function of Fe, Co and Ni retrieved from REELS data (open circles) compared with  Palik's optical data (dashed
curves). For the volume loss function of Fe and Ni, Tanuma's data are also shown as solid curves. (a) Fe; (b) Co; (c)
 Ni.}
\label{fdd}
\end{figure}

The surface (upper panels) and bulk (lower panels) loss functions of Palik's data and the present results are compared
in Figure~\ref{fdd}. For Fe and Ni, the data for the bulk loss function used by Tanuma, Powell and Penn \cite{tansia21}
is also shown for comparison. The general trend observed in these results is that for energies  above $\sim$15~eV, the
present data show more detailed structure in the loss function, while for lower energies, the earlier data seem to be
more realistic. This is again attributable to the limited energy resolution used in the present study and the
resulting problem with the elimination of the elastic peak.

To subject the present optical data to the usual sum--rule checks, the bulk loss function was extended above 80~eV by
Palik's data. The results for the perfect screening (or "ps") sum rule and the Thomas-Reiche-Kuhn (or "f") sum rule are
compared with the corresponding results based on Palik'sâ and Tanuma's loss functions in
Table~2. The ps-sum rule seems to  be the most important for the present study, since it emphasizes low energies,
while the main contribution to the f-sum rule comes from the core electrons which are not fully included in the present
measurements.  Except for Ni, the ps-sum rule check for the REELS data is in better agreement with the expected value of
1.000 than for the two other sets of optical data. The f-sum rule shows deviations of the expected value of
atomic electrons of the same order of magnitude for all data sets.

\begin{figure}[htb]
{\includegraphics[width=11.0cm]{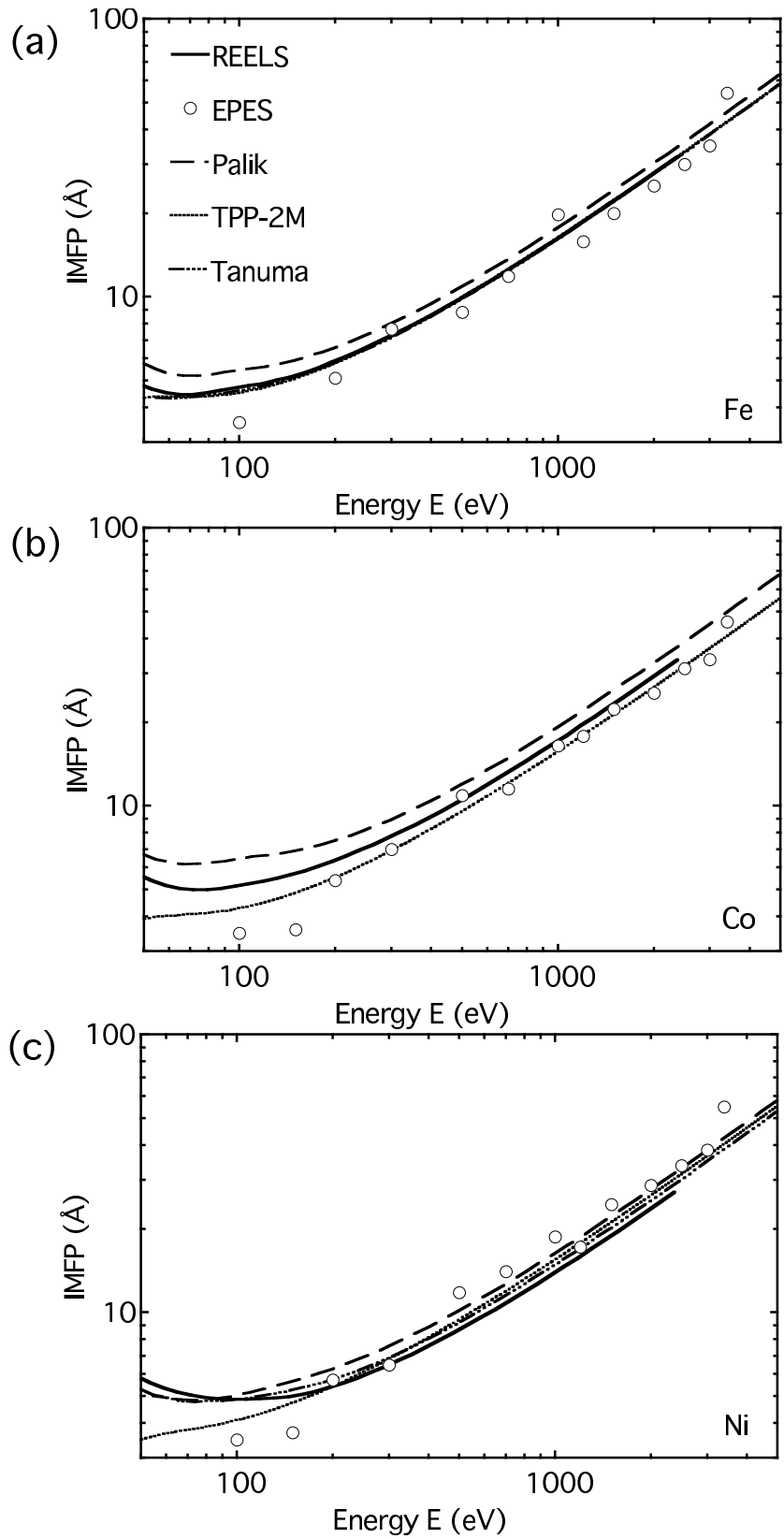}}
\caption{%
Inelastic mean free path of Fe, Co and Ni calculated from  the optical data shown in Figure~\ref{fdd} (solid curves);
Open circles represent experimental results measured with elastic peak electron spectroscopy (EPES)
\cite{werepesssl,werepesjes}; Dashed curves represent calculations based on  Palik's optical data \cite{palik,palik1};
The dotted curves represents the values predicted by the  TPP-2M formula \cite{tansia21};
Chain dashed curves for Fe and Ni represent calculations based on  Tanuma's optical data \cite{tanpriv} (a) Fe; (b) Co;
(c)
 Ni.
}
\label{fimfp}
\end{figure}

 Figure~\ref{fimfp} shows the IMFP for the studied materials over the energy range between 50 and 5000~eV as thick solid
line. For comparison, results using the other two sets of optical data are  also shown as dashed (Palik) and chain
dashed (Tanuma)  curves. For the calculation of the IMFP of all three sets of optical data the software employed by the
authors of Ref.~\cite{tansia21} was used \cite{tanpriv}. The results predicted by the TPP-2M-formula are indicated by
the dotted line. The open circles represent the results of elastic peak electron spectroscopy (EPES) measurements
reported earlier \cite{werepesjes,werepesssl}. The deviations of the IMFP values based on Palik'sâ data set and the
present ones are most prominent for Co, which is caused by the lower value of Palik's data for the loss function in the
region between 20 and 50~eV. For Fe and Ni, the mutual agreement between the different IMFP values is satisfactory,
except for energies below 100~eV, where the semiempirical TPP-2M formula predicts values for Co and Ni that are slightly
lower than the other results. The EPES data differ from the other data sets in that the experimental elastic peak
intensities are interpreted using only the elastic scattering cross section as input to the evaluation procedure
\cite{werepesjes,powjabepes}, while the other calculations all are based on optical data and dielectric response theory.
The two approaches are thus in a way fundamentally different. Nonetheless, the agreement between the EPES data and the
IMFP values derived from the dielectric function is reasonable, at least within the experimental scatter of the EPES
data.

\begin{figure}[htb]
{\includegraphics[width=11.0cm]{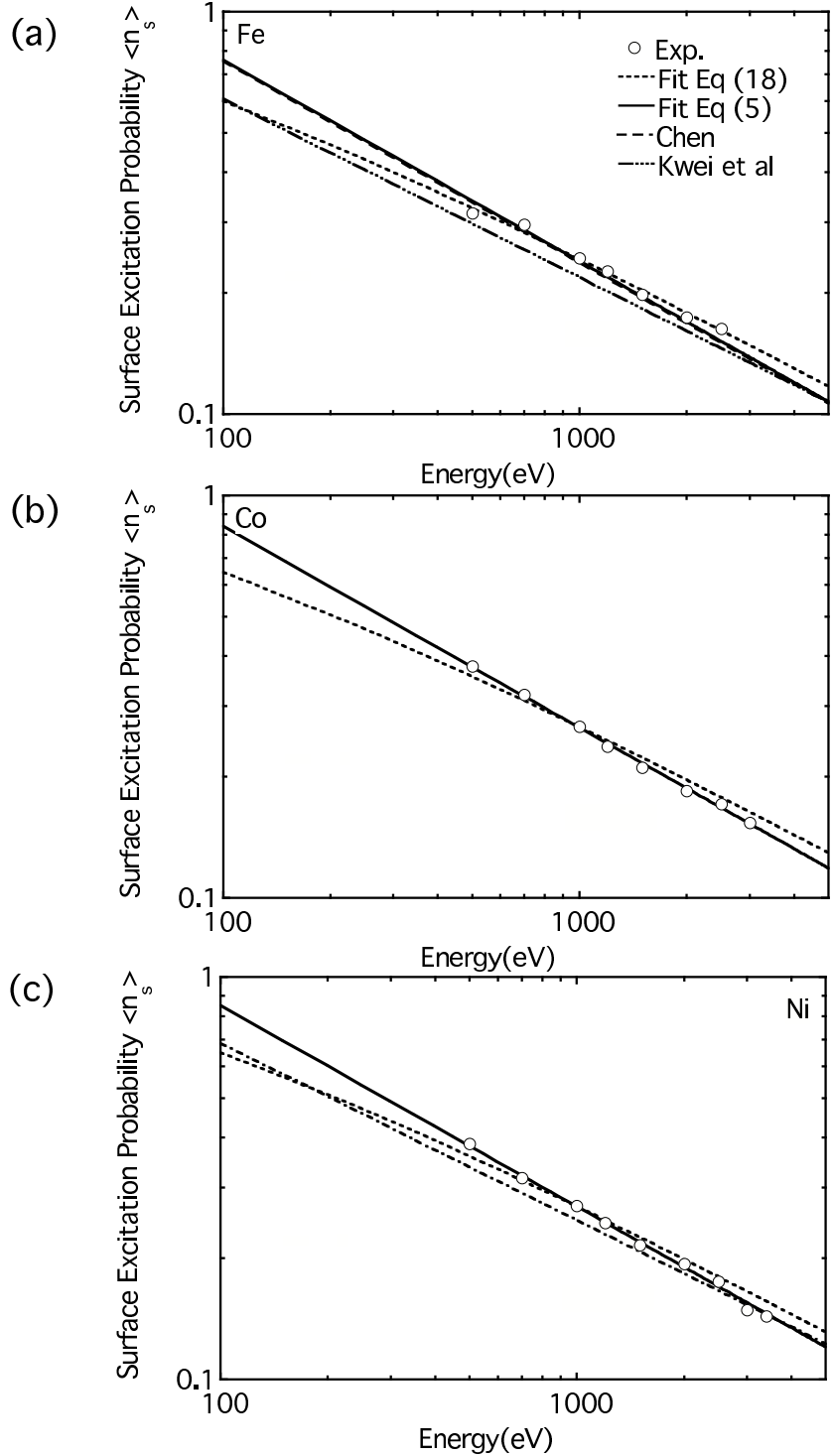}}
\caption{%
Surface excitation probability as a function of the energy for (a) Fe; (b) Co;  and (c) Ni (open circles). 
The solid and dashed lines are  fits of these data  to Eqn.~(\ref{ens} and Eqn.~(\ref{eoswald}) respectively). 
The dashed line shown for Fe is the theoretical dependence by Chen \cite{chenss519} which is almost impossible to
distinguish from the present data. The chain-dashed curves for Fe and Ni represent the results by Kwei et al \cite{kweisia26}.
}
\label{fsux}
\end{figure}

As a final result, Figure~\ref{fsux} shows the surface excitation probability extracted from the REELS spectra as open
circles. This quantity  was determined by using the present optical data to calculate the normalized  DIIMFP and DSEP
and by fitting the experimental spectrum to theory via Eqn.~(\ref{ereels}) using  $\langle n_{s} \rangle$ and the bulk
partial intensities as fit parameters. Examples of such fits are shown in Figure~\ref{freels} as thick solid lines. 
These fits are generally better than in previous work, where Palik's optica data were used for the same purpose
\cite{wersuxssl}. The contribution of electrons that have experienced one bulk, one surface and two surface excitations
is also indicated in these figures. The solid line  in Figure~\ref{fsux} is a fit of the data for the surface excitation
probability to Eqn.~(\ref{ens}), the dotted line represents a fit to another functional from of the surface excitation
probability that is commonly used \cite{oswphd}:
\begin{equation}
\label{eoswald}
\langle n_{s}\rangle=\frac{1}{a^*_{s}\sqrt{E}\mu_{i}+1}+\frac{1}{a^*_{s}\sqrt{E}\mu_{o}+1},
\end{equation}
the dashed and chain-dashed line are results by Chen\cite{chenss519} and Kwei et al \cite{kweisia26} respectively. The
quality of these fits is also improved compared to earlier work, in that the values of the surface excitation
probability $\langle n_{s} \rangle$ exhibit significantly less scatter than earlier \cite{wersuxssl}. The resulting
values of the surface excitation parameters $a_{s}$ and $a_{s}^*$ are given in Table 3. The present results for Fe are
in excellent agreement with the value of $a_{s}=2.51$, reported by Chen (dashed line in Figure~\ref{fsux}), while they
are also in close agreement with the results quoted by Kwei and coworkers \cite{kweisia26}. The quality of the fit of
the data to Eqn.~(\ref{ens}) is generally slightly better than the fit to Eqn.~(\ref{eoswald}), at least for Co and Ni,
although it is still difficult on the basis of the present data to decide between the two functional forms for the SEP,
Eqn.~(\ref{ens}) and Eqn.~(\ref{eoswald}). For this purpose, analysis of REELS experiments at higher energies would be
required.

\section{Summary and Conclusions.}
An earlier proposed procedure \cite{werbivar,werbivarss} for the simultaneous deconvolution of two REELS spectra to
provide the energy loss distribution in a single surface and volume electronic excitation is simplified using a Pad\'e
approximation and applied to experimental data of polycrystalline Fe, Co and Ni samples. It is shown that a second order
rational fraction approximation converges better over an energy loss range of about 150~eV than an 8-th order polynomial
expansion approximation, allowing one to conclude that the second order SW-algorithm can be safely used for practical
purposes. Analysis of the expansion coefficients provide guidelines on the choice of the optimal experimental parameters
to derive the DIIMFP and DSEP from REELS spectra and show, moreover, that no prior information on surface excitations is
needed to perform the deconvolution. The retrieved DIIMFP and DSEP were fitted to the corresponding theoretical 
expressions giving the optical data of the studied solids in terms of a set of Drude-Lindhard parameters. Agreement
between the resulting  optical data as well as the IMFP derived from them with values based on optical data reported
earlier is quite good, supporting the validity of the procedure. The values for the surface excitation probability
retrieved from the data using the optical constants derived in this work are believed to be more reliable than the
values reported earlier \cite{wersuxssl} and indicate that the energy and angular dependence of the surface excitation
probability are described by Eqn.~(\ref{ens}) rather than by Eqn.~(\ref{eoswald}). The values for the surface
excitation parameter are in reasonable agreement with theoretical values.

\section{Acknowledgment}
The author is grateful to Dr. S. Tanuma for making his optical data and his computer code for calculation of the IMFP
available for the present comparison. Financial support  of the present work by the Austrian Science Foundation FWF 
through Project No. P15938-N02 is gratefully acknowledged

\section{Appendix: First and Second Order Pad\'e rational fraction expansion coefficients.}
The explicit expressions for the first order bulk expansion coefficients in Eqn.~(\ref{eret}) in terms of the partial
intensities $\alpha_{n_{b}n_{s}}$ and $\beta_{n_{b}n_{s}}$ of two experimental  REELS spectra are given by:
\begin{eqnarray}
\label{epad1coef}
a^b_{00}&=&0
\nonumber\\
 a^b_{10}&=& {\beta_{01}}/\Delta 
\nonumber\\
 a^b_{01}&=& {-\alpha_{01}}/\Delta 
\nonumber\\
b^b_{00}&=&1
\nonumber\\
 b^b_{10}&=&
 \frac{1}{\Delta^2}\Big\{
 \alpha_{20} \beta_{01}^2-\alpha_{11} \beta_{01} \beta_{10}
+\alpha_{02} \beta_{10}^2
+\alpha_{01} \beta_{10} \beta_{11}
-\alpha_{01} \beta_{01} \beta_{20}
-\frac{\alpha_{01} \beta_{02} \beta_{10}^2}{\beta_{01}}\Big\}
\nonumber\\
 b^b_{01}&=&
 \frac{1}{\Delta^2}\Big\{
\alpha_{10}^2 \beta_{02}
-\alpha_{01} \alpha_{10} \beta_{11}
+\alpha_{01}^2 \beta_{20}
+\alpha_{10} \alpha_{11} \beta_{01}
-\alpha_{01} \alpha_{20} \beta_{01}
-\frac{\alpha_{02} \alpha_{10}^2 \beta_{01}}{\alpha_{01}}\Big\}
\end{eqnarray}
where the determinant $\Delta$ is given by
\begin{equation}
\label{edet}
\Delta={(\alpha_{10}\beta_{01} -\alpha_{01}\beta_{10})}
\end{equation}
The surface expansion coefficients read:
\begin{eqnarray}
\label{epad2coeff}
a^s_{00}&=&0
\nonumber\\
a^s_{10}&=& {-\beta_{10}}/\Delta
\nonumber\\
 a^s_{01}&=& {\alpha_{10}}/\Delta
\nonumber\\
b^s_{00}&=&1
\nonumber\\
 b^s_{10}&=& \frac{1}{\Delta^2}\Big\{
\alpha_{20} \beta_{01}^2-\alpha_{11} \beta_{01} \beta_{10}-\alpha_{10} \beta_{02} \beta_{10}+\alpha_{02} \beta_{10}^2+
\alpha_{10} \beta_{01} \beta_{11}-
\frac{\alpha_{10} \beta_{01}^2 \beta_{20}}{\beta_{10}}
\Big\}
\nonumber\\
 b^s_{01}&=& \frac{1}{\Delta^2}\Big\{
\alpha_{10}^2 \beta_{02}
-\alpha_{01} \alpha_{10} \beta_{11}
-\alpha_{02} \alpha_{10} \beta_{10}
+\alpha_{01}^2 \beta_{20}
+\alpha_{01} \alpha_{11} \beta_{10}
-\frac{\alpha_{01}^2 \alpha_{20} \beta_{10}}{\alpha_{10}}
\Big\}
\end{eqnarray}

The expressions for the second order expansion coefficients are somewhat too lengthy to reproduce here. They can be
easily calculated numerically though, by first establishing the polynomial expansion  coefficients $u^b_{pq}$ and
$u^s_{pq}$, as described in Refs.\cite{werbivar,werbivarss} up to third order ($p+q\le 3$). The coefficients in the
denominator of the rational fraction expansion are then found by solution of the homogeneous system of equations: 
 \begin{eqnarray}
\label{epade2homo}
0&=&b_{00}\nonumber\\
0&=&u_{30}+b_{10}u_{20}+b_{20}u_{10}\nonumber\\
0&=&u_{21}+b_{10}u_{11}+b_{01}u_{20}+b_{11}u_{10}+b_{20}u_{10}\nonumber\\
0&=&u_{12}+b_{10}u_{02}+b_{01}u_{11}+b_{11}u_{01}+b_{02}u_{10}\nonumber\\
0&=&u_{03}+b_{01}u_{02}+b_{02}u_{01}\nonumber\\
0&=&u_{22}+b_{10}u_{12}+b_{01}u_{21}+b_{11}u_{11}+b_{20}u_{02}+b_{02}u_{20}.
\end{eqnarray}
Finally, the enumerator coefficients are determined by the inhomogeneous system:
\begin{eqnarray}
\label{epade2inhomo}
a_{00}&=&u_{00}=0\nonumber\\
a_{10}&=&u_{10}\nonumber\\
a_{01}&=&u_{01}\nonumber\\
a_{11}&=&u_{11}+b_{10}u_{01}+b_{01}u_{10}\nonumber\\
a_{20}&=&u_{20}+b_{10}u_{10}\nonumber\\
a_{02}&=&u_{02}+b_{01}u_{01}.
\end{eqnarray}
\newpage


\newpage
\begin{table}
\begin{center}
\label{tdrd}
\begin{tabular}{||c|c|c||c|c|c||c|c|c||c|c|c||}
\hline
\hline
& Fe & 
&
& Co && Ni && \\
& $E(3p_{1/2})$ & &
& $E(3p_{1/2})$ & 
& $E(3p_{3/2})$ & &
\\
& $52.7~eV$ & &
& $58.9~eV$ & 
& $66.2~eV$ & &
\\
\hline
\hline
A${_i}$ (eV$^{2})$&           $\gamma_i$ (eV)&              $\omega_i$ (eV)&
A${_i}$ (eV$^{2})$&           $\gamma_i$ (eV)&              $\omega_i$ (eV)&
A${_i}$ (eV$^{2})$&           $\gamma_i$ (eV)&              $\omega_i$ (eV)\\
\hline
\hline
  0.39 & 7.54 & 0.0 & 10.89 & 8.16 & 0.0 & 79.78 & 8.16 & 0.0\\
  72.41 & 0.50 & 1.8 & 16.52 & 0.50 & 2.9 & 328.65 & 16.19 & 1.0\\
  170.71 & 5.95 & 4.2 & 10.89 & 0.50 & 4.2 & 76.99 & 15.64 & 2.5\\
  161.04 & 9.53 & 10.6 & 10.89 & 0.50 & 5.0 & 76.99 & 17.07 & 3.9\\
  91.21 & 9.12 & 19.1 & 49.02 & 2.51 & 6.2 & 63.58 & 7.04 & 14.5\\
  40.94 & 9.74 & 27.9 & 156.05 & 8.15 & 9.5 & 68.19 & 8.39 & 23.3\\
  24.17 & 4.78 & 34.9 & 99.00 & 13.94 & 15.0 & 305.75 & 39.04 & 41.9\\
  72.99 & 3.70 & 55.3 & 146.69 & 13.57 & 19.9 & 84.22 & 14.48 & 49.8\\
  64.36 & 8.06 & 60.8 & 10.89 & 3.10 & 25.5 & 164.52 & 29.48 & 64.2\\
  292.44 & 31.28 & 74.0 & 99.00 & 26.18 & 35.7 & 98.01 & 6.92 & 68.7\\
   &  &  & 15.39 & 5.62 & 38.8 & 7.59 & 1.35 & 76.9\\
   &  &  & 10.89 & 5.74 & 50.8 &  &  & \\
   &  &  & 60.53 & 3.92 & 60.9 &  &  & \\
   &  &  & 54.20 & 8.11 & 68.3 &  &  & \\
\hline
\hline
\end{tabular}
\caption{%
 Drude-Lindhard parameters for Fe, Co and Ni determined in the present work}
\end{center}
\end{table}

\begin{table}
\begin{center}
\label{tsr}
\begin{tabular}{||c||c|c||c|c||c|c||}
\hline
\hline
& Fe &(Z=26) &Co&(Z=27)&Ni&(Z=28)\\
\hline
Ref.& f-sum &ps-sum& f-sum &ps-sum& f-sum &ps-sum\\
\hline
\hline
REELS&24.0&1.007&24.2&0.925&29.4&1.030\\
Palik&24.0&0.943&24.2&0.845&26.7&1.010\\
Tanuma&23.5&1.110&--&--&27.3&1.050\\
\hline
\hline
\end{tabular}
\caption{%
Comparison of f- and ps-sum rule checks for different sets of optical data. The REELS data above 100~eV were extended to
30~keV using the data set of Palik }
\end{center}
\end{table}

\begin{table}
\begin{center}
\label{tsep}
\begin{tabular}{||c||c|c||}
\hline
\hline
& $a_{s}$ (Eqn.~(\ref{ens}))&$a_{s}^*$(Eqn.~(\ref{eoswald}))\\
& $(eV^{1/2})$&$(eV^{-1/2})$\\
\hline
\hline
Fe&2.53&0.34\\
Co&2.79&0.30\\
Ni&2.84&0.30\\
\hline
\hline
\end{tabular}
\caption{%
Surface excitation parameter $a_{s}$, according to Eqn.~(\ref{ens}) and $a_{s}^*$, according to Eqn.~(\ref{eoswald})
determined from the surface excitation probabilities shown in Figure~\ref{fsux}  }
\end{center}
\end{table}
\newpage

\newpage



\end{document}